\documentclass[twocolumn,aps,floats,superscriptaddress]{revtex4}
\usepackage{amsmath}
\usepackage{amssymb}
\usepackage{graphicx}

\newcommand{\mr}[1]{{{\mathrm{#1}}}}
\newcommand{\mcal}[1]{{\mathcal{#1}}}
\newcommand{\dt}{\partial_\tau}
\newcommand{\inte}{\int_0^\beta \!\!\!\! \mr{d}\tau}
\newcommand{\w}{\omega}
\newcommand{\pdagger}{{\phantom{\dagger}}}
\newcommand{\fasd}{f^\dagger_{\alpha\sigma}}
\newcommand{\fas}{f^\pdagger_{\alpha\sigma}}
\newcommand{\cksd}{c^\dagger_{k\sigma}}
\newcommand{\cks}{c^\pdagger_{k\sigma}}
\newcommand{\csd}{c^\dagger_{\sigma}}
\newcommand{\cs}{c^\pdagger_{\sigma}}
\newcommand{\Bad}{B^\dagger_{\alpha}}
\newcommand{\Ba}{B^\pdagger_{\alpha}}

\newcommand{\fjas}{f^\pdagger_{j\alpha\sigma}}
\newcommand{\cjsd}{c^\dagger_{j\sigma}}

\newcommand{\Bjad}{B^\dagger_{j\alpha}}
\newcommand{\Bja}{B^\pdagger_{j\alpha}}
\newcommand{\s}{\sigma}

\begin{document}

\title{Singular dynamics and pseudogap formation in the underscreened Kondo 
impurity and Kondo lattice models}

\author{S. Florens}
\affiliation{\mbox{Institut f\"ur Theorie der Kondensierten Materie, Universit\"at
Karlsruhe, 76128 Karlsruhe, Germany}}

\begin{abstract} 
\vspace{0.5cm} 
We study a generalization of the Kondo model in which the impurity spin is represented by 
Abrikosov fermions in a rotation group SU($P$) {\it larger} than the SU($N$) group
associated to the spin of the conduction electrons,
thereby forcing the single electronic bath to underscreen the localized moment. We
demonstrate how to formulate a controlled large $N$ limit preserving the
property of underscreening, and which can be seen as a ``dual'' theory of the
multichannel large $N$ equations usually associated to overscreening. 
Due to the anomalous scattering on the uncompensated degrees of
freedom, the Fermi liquid description of the electronic
fluid is invalidated, with the logarithmic singularities known to
occur in the $S=1$ SU(2) Kondo impurity model being replaced by continuous 
power laws at $N=\infty$.
The present technique can be extended to tackle the related underscreened Kondo 
{\it lattice} model in the large $N$ limit. We discover the occurence of an
{\it insulating pseudogap regime} in place of the expected renormalized metallic 
phase of the fully screened case, preventing the establishement of coherence 
over the lattice. This work and the recent observation of a similar weakly
insulating behavior on transport in CeCuAs$_2$ should give momentum for further 
studies of underscreened impurity models on the lattice.

\end{abstract}

\maketitle

\section{Introduction}

The Kondo model, initially introduced to describe with great success the
behavior of diluted magnetic impurities in metals~\cite{Hewson}, has been an 
exciting playground for condensed matter physicists in the last two decades. 
Indeed, the extreme simplicity of this many-body Hamiltonian was key to some 
unexpected progresses in the field, such as the discovery of the effective theory 
describing the Mott transition \cite{RMP_DMFT}, and as a prototypal description of 
Non Fermi liquid behavior in multi-channel \cite{Noz_Bland} and pseudogap extensions 
\cite{Bulla_Vojta}. Localized quantum impurities also provide the building entity 
to describe heavy fermion materials \cite{revue_HF}, and this is still a strong 
motivation for studying Kondo models on the lattice, most interestingly in 
relation to the violation of Fermi liquid behavior associated to the breakdown of Kondo 
screening close to a quantum critical point \cite{Stewart}, as discovered in a wide 
class of $f$-electron metals such as CeCu$_{6-x}$Au$_x$ \cite{schroeder}.
Interest in the Kondo model was also revived by the possibility of building and
controlling precisely quantum impurity states in semiconducting devices, leading to
the realization of the Kondo effect in quantum dots~\cite{revival_kondo}. The
possibility of observing Non Fermi liquid behavior in a multi-channel setup was 
recently advocated \cite{Oreg_GG,SF_AR,anders}, and provides further incentive for
a complete understanding of quantum impurity problems.

Quite recently, peculiar attention was given to the Kondo problem by Coleman and
Pepin \cite{Coleman_Pepin}, in the study of an underscreened model which
showed relatively strong deviations from Fermi liquid behavior in the specific
heat (we point out however that related effects can be traced back to the
physics of the ferromagnetic Kondo model~\cite{Giam_Varma}, see also~\cite{UKM_BA}).
This surprising result was interpreted as due to anomalous
scattering of conduction electrons onto the remaining unscreened spin degrees of
freedom, giving rise to a ``singular'' Fermi liquid fixed point, which differs from the 
intermediate fixed point~\cite{Noz_Bland,SF_scaling} associated to usual Non Fermi liquid by the
fact that the effective ferromagnetic Kondo coupling renormalizes to zero at low energy~\cite{Noz_Bland}.
This insightful work was however limited to a range of frequencies higher 
than the Zeeman energy, and concentrated mainly on thermodynamic quantities. In order to
study in more detail the dynamics due to these anomalous excitations in the underscreened Kondo 
model, we would like to find a simple approach that would be able to grasp the full crossover 
from the local moment regime at high temperature down to the underscreened state
in which nearby conduction electrons are tightly bound to the impurity.
In the realistic case of SU(2) spins, we note that the deviation 
from exact screening can be in principle tuned \cite{Noz_Bland} by changing
the size $S$ of the impurity spin and/or the number of screening channels $M$,
obtaining a transition from underscreening at $2S>M$ to overscreening at $2S<M$
(with perfect screening at $2S=M$).

A simple route to capture this physics consists in establishing a large $N$ limit 
of the problem, generalizing the quantum spin to the SU($N$) group. A crucial step lies
however in the choice of the SU($N$) representation in which the spin is considered. 
It is known that fermionic representations at large $N$ only allow perfect screening when $M \ll N$
\cite{Coleman2} and overscreening otherwise \cite{Cox_Ruck_multi,OP_AG_multi2}.  Interestingly,
bosonic representations of SU($N$) preserve the distinction between ``small''
and ``large'' spin as found by Parcollet and Georges \cite{OP_AG_multi1}, and
allow to study the underscreened situation. However, this case presents some
pathologies: the T-matrix scales as $1/N$, and the singular behavior as found
in~\cite{Coleman_Pepin} seems to be absent of the solution~\cite{AG_priv}. 

The previous discussion illustrates the need for an alternative large $N$ limit to
describe the underscreened Kondo effect, but also gives momentum to the idea we
will pursue in the following. Indeed, we can understand that overscreening and underscreening 
are somewhat ``dual'' in the sense that, while the former situation is reached in the presence of
many screening channels, the latter case should be obtained by considering many ``spin channels''.
A simple way to formalize this is to strongly enlarge the symmetry group of 
the impurity spin, thus forcing underscreening by the bath of conduction 
electrons. In fact, if one considers a generalized Kondo model involving a {\it single bath} of SU($N$) 
electrons interacting with a localized SU($P$) spin, where $P$ is larger than $N$, one
expects underscreening, {\it independently} of the representation chosen for the
impurity. Because the $M$-channel large $N$ limit which describes 
overscreening is obtained for $M \propto N$ \cite{Cox_Ruck_multi,OP_AG_multi2}, we can guess
that a reasonable large $N$ limit of underscreening can be found when $P = K N$ 
($K \propto N$ being the number of ``spin channels''). We will show that the saddle-point
equations derived in this way at $N=\infty$ present a ``dual'' structure 
to the Non Crossing Approximation associated to the overscreened case 
\cite{bickers_RMP,Cox_Ruck_multi,OP_AG_multi2}, and reveal non Fermi liquid behavior with a continuously 
varying exponent parametrized by the ratio $\gamma \equiv P/N^2$. 
The lattice version of this generalized underscreened Kondo model is also straightforwardly 
solvable, due to the absence of generated RKKY interactions. 
Because singular scattering of delocalized electrons on the degenerate 
manifold of unscreened spin occurs, coherence cannot establish over the lattice,
in contrast to the exactly screened Kondo lattice \cite{Auerbach_Levin}. This
proves the existence of a generic insulating-like state with a pseudogap density of
states, parametrized by the same anomalous exponent unveiled in the single
impurity underscreened Kondo model. We conclude the paper by future
prospects regarding the theory of underscreened Kondo models, and by the
possible observation of their physical consequences in strongly correlated
materials, pointing out some similarities with the cerium based compound 
CeCuAs$_2$, on which transport measurements were recently reported.

The remainder of the paper is organized as follows: in section~\ref{sec:derivation} 
we demonstrate how the new large $N$ limit of the underscreened Kondo impurity 
model can be performed, followed by a physical discussion of the results in 
section~\ref{sec:study}. The Kondo lattice extension is examined in 
section~\ref{sec:lattice}.

\section{Novel large $N$ limit of the underscreened Kondo model}
\label{sec:derivation}

\subsection{Model}
Our aim in this section is to introduce a Kondo model in which the
underscreened aspect is built in from the beginning, while a simple
large $N$ limit of the problem can be found by proper rescaling of the 
parameters.

\subsubsection{Heuristic derivation}
The basic idea, which was presented in the introduction, is to consider a single-channel 
Kondo Hamiltonian involving conduction electrons with $N$ spin flavors and interacting 
with a SU($P$) spin $S_{mm'}$ localized at the origin:
\begin{equation}
H = \sum_{k\s} \epsilon_k \cksd \cks + \sum_{kk'\s\s'mm'} J_{\s\s'}^{mm'}
\cksd c^\pdagger_{k'\sigma'} \; S_{m'm}
\end{equation}
where $\s=1\ldots N$, $m=1\ldots P$. The matrix of coupling constants
$J_{\s\s'}^{mm'}$ will be specified in the following. We choose a completely 
antisymmetric representation of the spin using Abrikosov fermions:
\begin{eqnarray}
S_{m'm} & = & f^\dagger_{m'} f^\pdagger_{m} - \frac{Q}{P} \delta_{mm'}\\
\label{constraint}
Q & = & \sum_{m} f^\dagger_{m} f^\pdagger_{m}
\end{eqnarray}
where the second equation is the necessary constraint to enforce the spin size 
$Q \equiv q_0 P$ (this scaling, where $q_0$ is of order 1, is necessary to get 
a large $N$ limit, see below). 
As emphasized previously, $P > N$ leads to underscreening,
and a likely scaling to obtain a solvable limit is $P \propto N^2$. Let us
therefore set $P \equiv K N $, with $K = \gamma N$, expressing fermions in a
double index notation:
\begin{equation}
S_{m'm} = f^\dagger_{\alpha'\s'} \fas - \frac{Q}{P}\delta_{\alpha\alpha'}\delta_{\s\s'} 
\end{equation}
where $m$ represents the doublet of indices $(\alpha,\s)$, with $\alpha = 1 \ldots K$. 
Neglecting potential scattering terms, which only contribute to next order in
$1/N$, the Hamiltonian now reads:
\begin{equation}
H = \sum_{k\s} \epsilon_k \cksd \cks +\!\!\!\!\! \sum_{kk'\s\s'\alpha\alpha'}
\!\!\! J_{\s\s'}^{\alpha\alpha'\s\s'} \cksd c^\pdagger_{k'\sigma'} \;
\!\! f^\dagger_{\alpha'\s'} \fas .
\end{equation}
This is now completely general, and we would like to chose the simplest coupling
between the itinerant SU($N$) fermions and the localized SU($KN$) spin. In the spirit of
having $K$ ``spin channels'', we set:
\begin{equation}
J_{\s\s'}^{\alpha\alpha'\s\s'} = \frac{J}{N} \delta_{\alpha \alpha'}
\end{equation}
and obtain the Hamiltonian 
\begin{eqnarray}
\label{Hfinal}
H & = & \sum_{k\s} \epsilon_k \cksd \cks +\!\!\!\!\! \sum_{kk'\s\s'\alpha}
\!\!\! \frac{J}{N} \cksd c^\pdagger_{k'\sigma'} \;
\!\! f^\dagger_{\alpha\s'} \fas \\
\label{new_constraint}
q_0 P & = & \sum_{\alpha\s} f^\dagger_{\alpha\s} \fas 
\end{eqnarray}
associated with the {\it single} constraint eq.~(\ref{new_constraint}).
This will be our starting point for the large $N$ solution.

\subsubsection{Alternative interpretation}
\label{sec:alt}

One might argue that coupling a band of fermions to a spin with a different symmetry group
is not a very transparent concept on a physical point of view. Indeed, one can
try to view alternatively the Hamiltonian~(\ref{Hfinal}) as the standard SU($N$) Kondo
model with a localized SU($N$) spin taken in a {\it rectangular}
representation~\cite{kiselev} of size $K\times q_0 N$:
\begin{equation}
S_{\s\s'} = \sum_\alpha  f^\dagger_{\alpha\s} f^\pdagger_{\alpha\sigma'} .
\end{equation}
Usual Kondo coupling to the SU($N$) electronic bath leads indeed to the Hamiltonian~(\ref{Hfinal}).
We will however not pursue this route in the present work, as this type of representation
asks to consider a {\it set} of $N^2$ constraints:
\begin{equation}
q_0 N \delta_{\alpha\alpha'} = \sum_{\sigma}
f^\dagger_{\alpha\sigma} f^\pdagger_{\alpha'\sigma}
\end{equation}
which makes the large $N$ limit impracticable. We will come back to the
differences between the large group technique (developed in this work) and the 
rectangular representation approach in section~\ref{sec:comparison}, while discussing 
the physical results.

\subsection{Derivation of the saddle-point equations}

In order to solve the model~(\ref{Hfinal}-\ref{new_constraint}), we first write the imaginary time 
action of the problem with inverse temperature $\beta$:
\begin{eqnarray}
\label{action}
\nonumber
S & = & \inte \bigg[ \sum_{\alpha \s} \fasd (\dt + \lambda) \fas + 
\sum_{k\s} \cksd (\dt + \epsilon_k) \cks \bigg] \\
& + & \inte \; \bigg[ \frac{J}{N} \sum_{kk'\s\s'\alpha} \cksd c^\pdagger_{k'\sigma'} \;
 f^\dagger_{\alpha\s'} \fas - q_0 P \lambda \bigg] 
\end{eqnarray}
introducing a Lagrange multiplier $\lambda$ to enforce the constraint~(\ref{new_constraint}).
The precise form of the action~(\ref{action}) is hinting that a large $N$ solution 
is possible, which we perform now.

The following step is to decouple the interaction with a bosonic field $\Bad$:
\begin{eqnarray}
S & = & \!\! \inte \bigg[ \sum_{\alpha \s} \fasd (\dt + \lambda) \fas + 
\sum_{k\s} \cksd (\dt + \epsilon_k) \cks \bigg] \\
\nonumber
& + & \inte  \bigg[ \sum_\alpha \frac{\Bad \Ba}{J} - q_0 P \lambda
+ \sum_{k \alpha \s} \cksd \fas \frac{\Ba }{\sqrt{N}} + h.c. \bigg] 
\end{eqnarray}
and integrate out the fermions $\fasd$:
\begin{eqnarray}
S & = & \!\! \inte \bigg[ \sum_{k\s} \cksd (\dt + \epsilon_k) \cks 
+ \sum_\alpha \frac{\Bad \Ba}{J} - q_0 P \lambda \bigg] \\
\nonumber
& + & \inte \inte' \frac{1}{N} G_{f0}(\tau-\tau') \sum_{kk' \alpha \s} 
\big(\cksd \Ba\big)_{\tau} \big(\Bad c^\pdagger_{k'\sigma} \big)_{\tau'}
\end{eqnarray}
where $G_{f0}(i\w_n) = 1/(i\w_n - \lambda)$, with $\w_n = (2n+1)\pi/\beta$. 
We finally introduce the electron sitting at the impurity center, $\csd \equiv \sum_k
\cksd$, so that:
\begin{eqnarray}
S & = & \inte \bigg[  \sum_\alpha \frac{\Bad \Ba}{J} - q_0 P \lambda \bigg] \\
\nonumber
&-& \inte \!\! \inte' G_{c0}^{-1}(\tau\!-\!\tau') \sum_\s \csd(\tau) \cs(\tau') \\
\nonumber
& + & \inte \inte' \frac{1}{N} G_{f0}(\tau-\tau') \sum_{\alpha \s} 
\big(\csd \Ba\big)_{\tau} \big(\Bad c^\pdagger_{\sigma} \big)_{\tau'}
\end{eqnarray}
where $G_{c0}(i\w_n) = \sum_{k} 1/(i\w_n - \epsilon_k)$.
Because $K=\gamma N$ scales as $N$, the existence of a saddle-point is manifest
in the previous expression. Following~\cite{OP_AG_multi2}, we find the 
integral equations:
\begin{eqnarray}
\nonumber
G_c(i\w_n) &\!\equiv\!& \big<\csd(i\w_n) \cs(i\w_n)\big> 
= \frac{1}{G_{c0}^{-1}(i\w_n) - \Sigma_c(i\w_n)} \\
\label{Gc}
\\
\label{GB}
G_B(i\nu_n) &\!\equiv\!& \big<\Bad(i\nu_n) \Ba(i\nu_n)\big>  
= \frac{1}{1/J - \Sigma_B(i\nu_n)}\\
\label{Sc}
\Sigma_c(\tau) &\!=\!& \gamma G_{f0}(\tau) G_B(\tau) \\
\label{SB}
\Sigma_B(\tau) &\!=\!& G_{f0}(\beta-\tau) G_c(\tau) \\
\label{lambda}
q_0 &\!=\!& \big<\fasd(\tau=0^+) \fas(\tau=0)\big>
\end{eqnarray}
where $\nu_n = 2 n \pi/\beta$ is a bosonic Matsubara frequency.

\subsection{Interpretation of the formalism}

Let us first comment on the formal analogy that the system of 
equations~(\ref{Gc}-\ref{SB}) shares with the usual Non Crossing Approximation 
(NCA) \cite{bickers_RMP,Cox_Ruck_multi,OP_AG_multi2}. We notice indeed that the 
roles of the local bath electron $\csd$ and of the Abrikosov fermion $\fasd$ are 
simply exchanged with respect to the usual NCA structure, which, we recall, is
associated to overscreening \cite{OP_AG_multi2}. In particular, the self-energy now 
involves the propagator $G_{f0}(i\w_n) = 1/(i\w_n - \lambda)$ instead of
$G_{c0}(i\w_n) = \sum_{k} 1/(i\w_n - \epsilon_k)$ in the NCA, and this
difference will be shown below to radically affect the physics, with the occurence of
underscreening instead of overscreening. This interesting ``duality'' 
is one of the main results of the present work.

It is also very appealing to remark that the computation of physical observables
related to the itinerant band does not involve directly pseudo-fermion degrees of 
freedom, contrarily to the multichannel case. Indeed, the present theory is 
formulated directly in terms of the bath propagator, to which the T-matrix is
simply related by the relation:
\begin{eqnarray} 
G_c(k,k',i\w_n) & \equiv & \big< c^\dagger_{k'\sigma}(i\w_n) 
c^\pdagger_{k\sigma}(i\w_n) \big> \\
\nonumber
& = & \frac{\delta_{kk'}}{i\w_n - \epsilon_k} +
\frac{1}{i\w_n - \epsilon_k} T(i\w_n) \frac{1}{i\w_n - \epsilon_{k'}} \\
\end{eqnarray} 
If we sum the previous expression over all momenta, we find:
\begin{eqnarray}
T(i\w_n) & = & \frac{G_c(i\w_n)-G_{c0}(i\w_n)}{G_{c0}^2(i\w_n)} \\
\label{T_mat_Kondo}
& = & \frac{1}{1/\Sigma_c(i\w_n) - G_{c0}(i\w_n)}
\end{eqnarray}
where~(\ref{Gc}) was used to obtain the second expression.
Interestingly, we note that the T-matrix~(\ref{T_mat_Kondo}) is of order
$N^0$  in the present scheme, whereas it happens to scale as $1/N$ in the multi-channel large $N$ limit
\cite{OP_AG_multi2}, a certain drawback for the theory of overscreening.
However, by the ``duality'' argument, we expect to find that the spinon propagator
$G_f(i\w_n) \equiv \big<\fasd(i\w_n) \fas(i\w_n)\big> $ only shows $1/N$ corrections to the
free impurity limit ($J=0$), which can indeed be easily checked by a direct computation:
\begin{eqnarray}
G_f(\w_n) &=& G_{f0}(i\w_n) - G_{f0}^2(i\w_n) G_\chi(i\w_n) \\
G_\chi(\tau) & \equiv & \frac{1}{N} G_c(\tau) G_B(-\tau).
\end{eqnarray}
This means that, although the T-matrix is conveniently captured by the present
scheme, computation of e.g. the spin susceptibility involves necessarily $1/N$
corrections. This remark allows however to solve explicitely the constraint 
equation~(\ref{lambda}) at the leading order: 
\begin{eqnarray}
q_0 & = & G_{f0}(\tau=0^-) + \mcal{O}\big(\frac{1}{N}\big) 
= \frac{1}{e^{\beta\lambda}+1} \\
\label{lambda_dominant}
\Rightarrow \lambda & = & \frac{1}{\beta} \log \frac{1-q_0}{q_0} +
\mcal{O}\big(\frac{1}{N}\big).
\end{eqnarray}

\section{Physical study of the non Fermi liquid regime}
\label{sec:study}

\subsection{At particle-hole symmetry}
\begin{figure}
\begin{center}
\includegraphics[width=0.84 \linewidth]{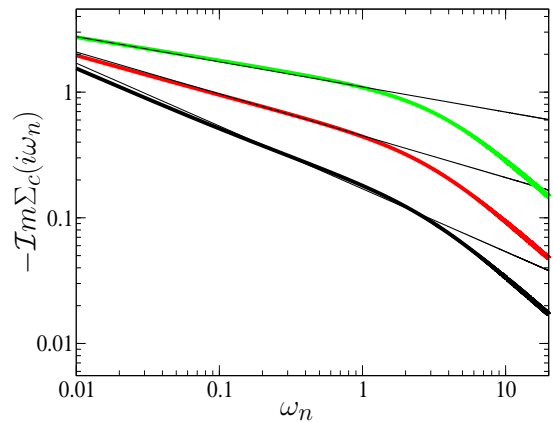}
\end{center}
\caption{(color online). Logarithmic plot of $-\mcal{I}m \Sigma_c(i\w_n)$ for $\gamma=1,3,10$
(bottom to top) at $\beta=1000$, $J=1$, $D=4$. Thin lines are fits to a power law
form $|\w|^{-\alpha}$ according to Eq.~(\ref{alpha}).}
\label{fig:power}
\end{figure}

For simplicity, we start with the assumption of particle-hole symmetry, 
$q_0 = 1/2$ (the free bath $G_{c0}(i\w_n)$ will always be assumed symmetric 
in the following). This implies that the constraint~(\ref{lambda}) is 
fulfilled provided $\lambda = 0$, and we have:
\begin{equation}
G_{f0}(\tau) = -\frac{1}{2} \mr{Sgn}(\tau)
\end{equation}
i.e. long range correlations in the self-energies~(\ref{Sc}-\ref{SB}).
By analogy with the low-frequency analysis usually performed in studying the NCA
equations \cite{mueller_hart,OP_AG_multi2}, we assume power law behavior of the
self-energies and plug into the saddle-point equations. Self-consistency can be
achieved at zero temperature (see Appendix~\ref{app1}) and we obtain for real 
frequency quantities:
\begin{eqnarray}
\label{Sigma_c}
\mcal{I}m \; \Sigma_c(\w) & = & C_c |\w|^{-\alpha} \\
\mcal{I}m \; \Sigma_B(\w) & = & C_B |\w|^{+\alpha} \mr{Sgn}(\w) \\
\label{alpha}
\alpha & = & \frac{2}{\pi} \arctan\frac{1}{\sqrt{\gamma}}
\end{eqnarray}
where $C_c$ and $C_b$ are undetermined constants.
This analytical result is well borne out by the numerical solution of the
saddle point equations as shown by figure~\ref{fig:power} (in the following
computations, a semi-circular density of states with half-width $D$ was chosen
to model the bath of conduction electrons).

We would like to contrast this peculiar low energy behavior with the one observed 
in a standard Fermi liquid (as in the fully screened Kondo model).
The interpretation of these results can be made clearer by considering
the T-matrix of a related Anderson model:
\begin{equation}
\label{T_mat_Anderson}
T(i\w_n) = \frac{V^2}{i\w_n - \Sigma_d(i\w_n) - V^2 G_{c0}(i\w_n)}
\end{equation}
where the local fermion $d^\dagger_{\sigma}$  has an hybridization $V$ to 
the $c^\dagger_\s$-electrons and is subject to a local Coulomb repulsion term
$U$ (and possibly to a strong Hund's rule if one wants to stabilize a $S=1$ spin
in order to make the connection to the underscreened case),
giving rise to the term $\Sigma_d(i\w_n)$ in the previous equation. 
Comparing~(\ref{T_mat_Kondo}) and~(\ref{T_mat_Anderson}), we see that 
the inverse self-energy of the bath electrons is simply
related to the impurity self-energy by $1/\Sigma_c(\w) = [\w - \Sigma_d(\w)]/V^2$. 
In a Fermi liquid, we know~\cite{Hewson} that the impurity self-energy obeys 
at low frequency:
\begin{equation}
\label{eq:SFL}
\Sigma_d^\mr{FL}(\w) = (1-1/\mcal{Z})\w + i A \w^2 +\ldots,
\end{equation}
where $\mcal{Z}$ is the quasiparticle residue and $A \w^2$ the inelastic scattering rate. 
The previous general identification between $1/\Sigma_c$ and $\Sigma_d$ provides the self-energy 
of the local electron in the bath for a Fermi liquid:
\begin{equation}
\Sigma_c^\mr{FL}(\w) = \frac{V^2}{\w-\Sigma_d^\mr{FL}(\w)} = 
\frac{\mcal{Z}V^2}{\w} + i A V^2 Z^2 + \ldots,
\end{equation}
the first term corresponding to regular elastic scattering from the impurity. The fact 
that this self-energy diverges as a power law at low frequency in the underscreened 
case~(\ref{Sigma_c}) signals anomalous scattering on the remaining unscreened degrees 
of freedom, which ultimately violates the Fermi liquid description of the problem. 
This result can be more clearly witnessed from the impurity self-energy itself
\begin{equation}
\Sigma_d(\w) = \w - \frac{V^2}{\Sigma_c(\w)} \propto |\w|^\alpha
\end{equation}
showing a power law scattering rate for the localized state which even dominates
the linear part found in the self-energy of a Fermi liquid,
equation~(\ref{eq:SFL}). Indeed, we deduce from this behavior a frequency dependent 
quasiparticle residue $\mcal{Z}(\w) = [1-\partial\Sigma_d(\w)/\partial\w]^{-1} 
\propto |\w|^{1-\alpha}$ which {\it vanishes} at low energy, demonstrating a clear
breakdown of Fermi liquid theory.

\begin{figure}
\begin{center}
\includegraphics[width=0.84 \linewidth]{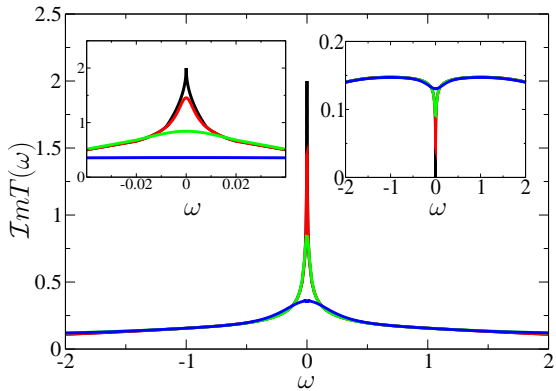}
\end{center}
\caption{(color online). Imaginary part of the T-matrix for decreasing temperatures
$\beta=10,100,1000,\infty$ 
(bottom to top) with $J=1$, $D=4$, $\gamma=1$. The left inset is a zoom on the
Kondo resonance, and the right inset shows the associated depletion of low energy
states in the local spectral function of the bath 
$(-1/\pi) \mcal{I}m G_c(\w) \propto |\w|^\alpha$.}
\label{fig:spectral}
\end{figure}

A final way to put into evidence the singular energy dependence of this
underscreened impurity is to compute the T-matrix. We find indeed that the
Kondo resonance, whose development is illustrated on figure~\ref{fig:spectral},
displays a cusp at low frequency:
\begin{equation}
- \mcal{I}m T(\w) = \frac{\pi \rho_0}{(\pi\rho_0)^2 + B |\w|^{2\alpha}}
\end{equation}
using in equation~(\ref{T_mat_Kondo}) the low frequency behavior of the free Green's function,
$G_{c0}(\w) = -i \pi \rho_0$. This expression shows that, although the unitary limit is 
recovered in the T-matrix at $\w=0$, Fermi liquid behavior is nevertheless violated by the
non-analytic expansion a low frequency.

\subsection{Away from particle-hole symmetry}
\label{phbreak}

Here we discuss briefly the case $q_0 \neq 1/2$, and investigate whether the
previous results remain valid when particle-hole symmetry is broken. The
propagator appearing in the bubble giving the self-energies,
equations~(\ref{Sc}-\ref{SB}), is given by:
\begin{equation}
G_{f0}(\tau) = - \frac{e^{-\lambda \tau}}{e^{-\lambda\beta}+1} \hspace{1cm}
\mr{for}  \;\; 0<\tau<\beta
\end{equation}
and decays exponentially. We argue that the long range correlations which are crucial
to maintain the non-trivial power laws are still present, because $\beta\lambda$
saturates at low temperature, from~(\ref{lambda_dominant}). 
This would imply that the non Fermi liquid state survives the introduction of 
the asymmetry parameter $q_0\neq 1/2$, as can also be verified from the numerics.

However, the structure of the saddle point is such that the constraint~(\ref{new_constraint})
scales as $N^2$ instead of $N$, and hints that it might be necessary to account
for $1/N$ corrections to the result~(\ref{lambda_dominant}). We do
not find that this possibility really modifies the previous result, $\lambda
\propto 1/\beta$, although we think this question deserves further clarification.

\subsection{Discussion: theoretical aspects}

\subsubsection{Possible extensions of the formalism}

The present approach seems well suited to describe the underscreened
situation of quantum impurity models. We can however point out two kinds of
limitations, which are in fact also inherent to the Non Crossing Approximation of
multichannel models as well~\cite{OP_AG_multi2}. First, although the T-matrix can now
be naturally extracted at the saddle-point level, some physical quantities such as
the spin susceptibility appear only at next to leading order in 1/$N$. Second,
motivated by some puzzles raised in heavy fermion compounds, it remains of great 
interest to develop faithfull large $N$ expansion methods that can work in the
exactly screened case too (an interesting step in this direction was however taken
in~\cite{OP_AG_multi1}). It was in fact already acknowledged~\cite{OP_these} that a
simple generalization of the Kondo model is quite suitable for overcoming both
of these difficulties: it consists in taking the localized spin in a rectangular
representation of SU($N$), as indeed done in the present work (if one were able
to take good care of the necessary constraints, see discussion in section~\ref{sec:alt}),
while assuming also a large number (of order $N$) of screening channels (in this
paper we only considered the single channel case). This extension of the formalism, 
also known as ``matrix Kondo model'', leads however to technical difficulties due to
proliferation of Feynman graphs at large $N$. It is actually possible to resum this 
complicated diagrammatics using a soft constraint approach inspired by a recent
study ~\cite{SF_AR} of the interplay of Kondo effect and Coulomb blockade 
(viewed now as a purely mathematical trick). Such a theory would be a bridge
between the usual large $N$ approach of overscreening and the present work,
both appearing as simple limitating cases of it, thus finding an unity behind our
``duality'' picture. Such a theory will be considered in a forthcoming publication~\cite{SF_todo}

\subsubsection{On the nature of the low energy singularities}
\label{sec:comparison}
As we have demonstrated from the analytical and numerical solution of the new large-N
equations, the underscreened Kondo impurity model displays violations of low
energy Fermi liquid behavior, although those are less severe as the one observed
in the overscreened case (where the deviation from the unitary limit associated 
to finite lifetime of the electrons at zero temperature even dominates the anomalous 
frequency dependence of the scattering rate). The underscreened
impurity is nevertheless associated to Non Fermi liquid behavior: the local quasiparticle 
weight vanishes at low energy and the system presents additional degeneracy due to the 
non fully screened spin. It is interesting to compare this result to previous exact 
knowledge from the Bethe ansatz solution of underscreened Kondo models. Although
the general features we have sketched so far are indeed obtained from the Bethe ansatz
analysis, one crucial difference concerns the precise nature of the singular
frequency dependence of the physical quantities: while we observe universal
non-trivial exponents in the large N solution, the $S=1$ SU(2) Kondo model
rather displays logarithmic behavior~\cite{Coleman_Pepin} (although power laws
are actually found in the anisotropic case~\cite{UKM_BA}). We think this
difference is a minor point, that we would naively like to attribute to special features 
associated to the large N generalization of the model. This point is actually
more subtle. Indeed , the hamiltonian~(\ref{Hfinal}) we have solved can be argued
(see section~\ref{sec:alt}) to be equivalent to a SU($N$) Kondo model with spin in
a rectangular representation, which is actually diagonalized by Bethe
ansatz~\cite{PZJ}, and found to display also logarithmic singularities! As
stressed previously, we have however enforced a different type of constraint in
the present approach, and we would ultimately pin point this difference to be the 
origin of the discrepancy in the low energy behavior as compared to the exact solution. 
This discussion would tend to illustrate in any case the fact that the physics can be 
sensitive in how the constraint is implemented, but we will reinforce our judgment that
the present method captures the essential features of underscreening.
As we will now demonstrate, the main interest of our approach is that it can
apply to problems where no exact solution is available, in particular in the many
impurities extension of the underscreened Kondo model.

\section{Pseudogap formation in the underscreened Kondo lattice}
\label{sec:lattice}
\subsection{Model and large $N$ solution}

We introduce here the lattice extension of the previous single impurity model,
which consists of a dense network of impurities carrying a SU($P$) spin on 
which a band of SU($N$) itinerant electrons scatter:
\begin{equation}
H = \sum_{k\s} (\epsilon_k-\mu) \cksd \cks + \frac{J}{N} \sum_{j\s\s'\alpha}
\cjsd c^\pdagger_{j\sigma'} \;
f^\dagger_{j\alpha\s'} \fjas 
\end{equation}
Here $j$ labels sites, with $\cksd = \sum_j \cjsd e^{i k R_j}$, $\mu$ is the 
chemical potential in the $c$-band and other conventions are similar to previously. 
This model is not very realistic in many respects, and to date lacks a physical 
realization in condensed matter systems (see however the ending discussion in
section~\ref{discussion}). On a purely theoretical point of view, we can raise
however two interesting questions. First, what is the behavior of the electronic
band below the Kondo temperature associated to underscreening? For this to make
sense, we need to argue that the magnetic processes involving the
remanent spin degrees can be ignored, for example due to frustration (maybe by
analogy to the compound  LiV$_2$O$_4$). Second, when the partially-screened
local moments start to order, and quench their macroscopic entropy, how does the system 
behaves?  In the following large $N$ analysis of this problem, we will see that
intersite magnetic correlations are absent, and that the first of these
questions can be answered.

The large $N$ limit is derived following the same steps performed in
section~\ref{sec:derivation}, introducing local bosons $\Bjad$ to decouple the Kondo
term on each site, and integrating the Abrikosov fermions:
\begin{eqnarray}
\label{action_lattice}
S & = & \inte \sum_{k\s} \cksd (\dt + \epsilon_k - \mu) \cks 
+ \sum_{j\alpha} \frac{\Bjad \Bja}{J} \\
\nonumber
& + & \inte \inte' \frac{1}{N} G_{f0}(\tau-\tau') \sum_{j \alpha \s} 
\big(\cjsd \Bja\big)_{\tau} \big(\Bjad c^\pdagger_{j\sigma} \big)_{\tau'}
\end{eqnarray}
We can simply read off this expression the final saddle point equations, which
are completely identical to~(\ref{GB}-\ref{SB}), except that the local bath
propagator now reads:
\begin{eqnarray}
G_c(i\w_n) & = &  \big<c_{i\s}^\dagger(i\w_n) c_{i\s}^{\phantom{\dagger}}(i\w_n)\big> \\
\label{Gc_latt}
& = & \sum_k \frac{1}{i\w_n - \epsilon_k + \mu - \Sigma_c(i\w_n)}
\end{eqnarray}

\subsection{Interpretation}

The new system of integral equations~(\ref{GB}-\ref{SB}),(\ref{Gc_latt})
is remarkably simple in its structure, a fact due to the absence of
intersite correlations at this level of approximation. Indeed, as the
action~(\ref{action_lattice}) shows, no magnetic RKKY interaction is generated
at the saddle-point level, which is expected since the additional quantum number $\alpha$ 
carried by the localized spins cannot be transported from site to site by the itinerant 
fermions. We point out however that this is not a general feature of
underscreened models, and one can easily check that the RKKY interaction is
indeed mediated by the electronic bath in the finite $S$ SU(2) case.

Besides, expression~(\ref{Gc_latt}) is reminiscent of a self-consistent T-matrix
approximation and signals that electrons in the bath are still anomalously 
scattered by the localized spins, acting independently of each other. Indeed, 
if we assume that $\Sigma_c(i\w_n)$ is divergent as in the single 
impurity case, equation~(\ref{Sigma_c}), we see that this singular self-energy
dominates the $k$-summation in~(\ref{Gc_latt}) at low energy, so that:
\begin{equation}
\label{G_c_lattice}
G_c(i\w_n) \sim \frac{1}{-\Sigma_c(i\w_n)} \sim i |\w_n|^\alpha \mr{Sgn}(\w_n)
\end{equation}
with the {\it same} exponent~(\ref{alpha}) as found previously. This means that in the
underscreened Kondo lattice, there is no important distinction between the cases 
of dense and diluted impurities (on the point of view of the bath electrons that 
feel the Kondo interaction), and that the itinerant electron density of states always shows a
pseudogap at low energy. This is quite different from the situation of exactly 
screened models, where a hard hybridization gap would open at half-filling. 
In the perfectly screened case, coherence can always be re-established upon doping, 
despite the fact that each individual impurity scatters strongly the electrons. 
The result~(\ref{G_c_lattice}) would however let us think that electronic degrees 
of freedom always remain confined in the underscreened Kondo lattice, as we now check 
on the numerical solution of the saddle-point equations.

\subsection{Results}

\begin{figure}
\begin{center}
\includegraphics[width=0.84 \linewidth]{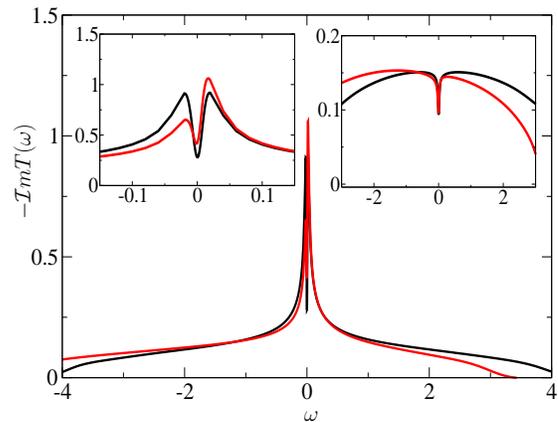}
\end{center}
\caption{(color online). Imaginary part of the T-matrix at $\beta = 100$ and for two values of the
chemical potential $\mu=0$ (symmetric curve) and $\mu=1$ (asymmetric curve), 
with $J=1$, $D=4$, $\gamma=1$, showing the presence of an hybridization pseudogap (see 
zoom in left inset). The right inset shows the corresponding dip in $(-1/\pi)\mcal{I}m G_c(\w)$. 
Because of finite temperature effects, spectral weight is present at zero frequency.}
\label{fig:lattice}
\end{figure}

We will be again interested in the T-matrix, which is related to the
(translation invariant) $c$-electron Green's function $G_c(k,i\w_n) 
\equiv \big< c^\dagger_{k\sigma}(i\w_n) c^\pdagger_{k\sigma}(i\w_n) \big>$ by:
\begin{equation}
G_c(k,i\w_n) = \frac{1}{i\w_n - \epsilon_k + \mu} +
\frac{T(k,i\w_n)}{(i\w_n - \epsilon_k+\mu)^2} 
\end{equation}
This relation can be best understood from an equivalent periodic Anderson model, where
$T(k,i\w_n)$ is proportional to the momentum- and frequency-dependent Green's 
function of the localized electrons. From the effective
action~(\ref{action_lattice}) we have:
\begin{equation}
\label{local_self}
G_c(k,i\w_n) = \frac{1}{i\w_n - \epsilon_k + \mu - \Sigma_c(i\w_n)}
\end{equation}
so that the T-matrix can be expressed as:
\begin{equation}
\label{T_lattice}
T(k,i\w_n) = \frac{1}{1/\Sigma_c(i\w_n)-1/(i\w_n - \epsilon_k + \mu)}
\end{equation}
Again, we identify $1/\Sigma_c(i\w_n)$ as the impurity self-energy
$\Sigma_d(i\w_n)$ (up to a factor $V^2$).
The local T-matrix, $T(i\w_n) \equiv \sum_k T(k,i\w_n)$ is easily calculated 
from~(\ref{T_lattice}) after the numerical solution of the saddle-point
equations, and is shown in Figure~\ref{fig:lattice} (here also a semi-circular
density of states for the $c$-electron was taken).

From the numerical solution, and in agreement with the previous analytical
analysis, we see that an {\it hybridization pseudogap} opens in the spectrum,
irrespective of the filling of the $c$-band. This prevents coherence to be 
reached over the lattice at zero temperature (note that Figure~\ref{fig:lattice} 
is performed at finite temperature, so that the pseudogap in the density of
states is filled by thermal excitations). Therefore, the strongly underscreened Kondo 
lattice is strictly speaking an insulator, with a resistivity likely showing a power law 
increase as temperature is lowered, instead of the activated or metallic behavior 
expected respectively in usual Kondo insulators or in heavy fermion metals.
In order to shed some light onto this remarkable result, we put it into the
perspective of previous theoretical work and experimental measurements on a newly
discovered $f$-electron material.

\subsection{Physical discussion}
\label{discussion}

\subsubsection{Underscreened Kondo lattice: status of theory}

The main physical result of the paper is that the underscreened Kondo lattice
is an insulator with a pseudogap density of states above the ordering
temperature of the partially screened moments, for all filling of the electronic band. 
Theoretically, we can ask whether this behavior is peculiar to the large $N$ solution 
performed here or can be expected to be generic. To answer this question, it is interesting 
to notice two facts. First, previous work by K. Le Hur in a one dimensional model
of underscreened Kondo ladder~\cite{KLH_UKM} demonstrated infact similar insulating behavior 
(this quite peculiar model was chosen in order to prevent the Haldane spin gap 
associated to the 1D $S=1$ chain, which would pollute the interesting physics).
Second, our large $N$ solution displays actually some high dimensional
features~\cite{RMP_DMFT}, such as a local self-energy, as seen in equation~(\ref{local_self}).
Because this insulating behavior is found in two such extreme cases, we could expect for 
a continuity of this non-metallic behavior of underscreened lattice in intermediate dimension also.

We now turn to the nature of this insulating state. As discussed in the single
impurity case, section~\ref{sec:comparison}, the large $N$ solution predicts
that the dynamics is governed by universal power law behavior rather than the
logarithmic singularities found in the $S=1$ SU(2) case and extensions, although
the physics of these anomalies can be expected to be similar. This discussion
extends naturally to the underscreened lattice, and deserves clearly further
theoretical studies. We think that a next step could be taken by solving the
$S=1$ Kondo lattice model in a combination of DMFT~\cite{RMP_DMFT} and NRG techniques,
and would help both to confirm the generic insulating behavior found here, and
investigate in more detail the physical properties that are associated to it.

\subsubsection{Physical observation of the underscreened Kondo effect}

We finally discuss the possible experimental realization of underscreened Kondo
physics. Concerning the single impurity case, the stability of $S=1$ for quantum dots 
in even valleys of Coulomb blockade was discussed both theoretically and experimentally
(for a review, see~\cite{S1_kondo_dot}), although no emphasis was put onto the singular
behavior that we have discussed previously. Another way to observe Kondo
underscreening would be to study large $S$ atoms deposited on a metallic surface
and scanned by an STM tip.

Finding an $f$-electron material related to the underscreened Kondo lattice
seems to be a more daunting task. Indeed, not only the intersite magnetic 
correlations would need to be small in such a system, but a mechanism to form
large spin (or spin-orbital degeneracy) of the $f$-level should be invoked.
Typically, crystal fields leave unfortunately an effective Kramers doublet below
temperature of several dozens of Kelvins. However, we think that the
present work is interesting since it highlights the peculiar physical 
properties of underscreened lattices, and may be useful in strengthening an 
eventual experimental observation. In fact, very recent transport measurements
of a new cerium based compound, CeCuAs$_2$, revealed weakly insulating behavior of
the resistivity, with a power law divergence at low temperature~\cite{sampath}. 
While we are careful in connecting hastily this observation to our theoretical findinds, 
we hope that these remarks would stimulate further studies of underscreening
effects in Kondo systems.

\section{Conclusion}

The underscreened Kondo model was investigated in this paper by means of a
specially developed large $N$ technique. Although the strong coupling fixed
point in which itinerant electrons are tightly bound to the uncompensated spin
is stable in the renormalization group sense, we have shown that non Fermi liquid 
behavior appears in the form of anomalous power laws in the physical observables. 
The universal exponent was computed analytically and checked over the numerical 
solution of integral equations, which show an interesting duality with respect to 
the previous theory of overscreening in multichannel models.
The extension to a finite dimensional lattice of underscreened magnetic moments
was also considered, and a pseudogap (weakly insulating) regime was discovered, with 
similar power laws governing the dynamics of the electrons. We have also tried to
point out several directions for future research, both concerning the theoretical 
aspects of underscreened models and their experimental realization, for which 
CeCuAs$_2$ might be a candidate.

\acknowledgements
I wish to thank Matthias Vojta for some useful suggestions at the 
early stages of this work, and for comments on the manuscript. I am also
grateful to E. V. Sampathkumaran for communicating his experimental results
before publication~\cite{sampath}. Financial support by the Center for Functional Nanostrutures 
is also acknowledged. During the completion of the present paper, a similar 
large $N$ limit of the underscreened Kondo model was proposed by P. 
Coleman and I. Paul \cite{Coleman_Paul}, with results complementary to ours in
the single impurity case. A very detailed investigation of the singularities occuring 
in the $S=1$ SU(2) Kondo model was also carried out simultanenously by P. Mehta 
et al.~\cite{singular} using a combination of Bethe ansatz and Numerical Renormalization 
Group techniques.

\appendix
\section{}
\label{app1}

We present here the derivation of the non-trivial exponent governing the low
frequency behavior of the physical quantities, following 
\cite{OP_AG_multi2, mueller_hart}. Let us assume the ansatz:
\begin{eqnarray}
\mcal{I}m G_c(\w) &=& A_c |\w|^{-\alpha_c} \\
\mcal{I}m G_B(\w) &=& A_B |\w|^{-\alpha_B} \mr{Sgn}(\w)
\end{eqnarray}
at zero temperature. Using the spectral decomposition:
\begin{equation}
G_{c}(\tau) = \int_{0}^{+\infty}\frac{d\w}{\pi} \,\,
e^{-\w \tau} \, \mcal{I}m G_c(\w) 
\end{equation}
and the expression for the self-energies~(\ref{Sc}-\ref{SB}) with 
$G_{f0}(\tau) = -\frac{1}{2} \mr{Sgn}(\tau)$, we find:
\begin{eqnarray}
\Sigma_c(\tau) &=& -\frac{\gamma A_B}{2\pi} \Gamma(1-\alpha_B) 
\frac{\mr{Sgn}(\tau)}{|\tau|^{1-\alpha_B}} \\
\Sigma_B(\tau) &=& -\frac{A_c}{2\pi} \Gamma(1-\alpha_c) \frac{1}{|\tau|^{1-\alpha_c}}
\end{eqnarray}
Going back to frequency, we have simply:
\begin{eqnarray}
\mcal{I}m \Sigma_c(\w) &=& - \frac{\gamma A_B}{2} |\w|^{-\alpha_B} \\
\mcal{I}m \Sigma_B(\w) &=& - \frac{A_c}{2} |\w|^{-\alpha_c} \mr{Sgn}(\w)
\end{eqnarray}
To determine the real part in the previous self-energies, we use an analyticity
argument, which gives for complex frequency $z$:
\begin{eqnarray}
\Sigma_c(z) &=& - \frac{\gamma A_B}{2} 
\frac{e^{i(1+\alpha_B)\pi/2}}{\sin[(1+\alpha_B)\pi/2]} |z|^{-\alpha_B} \\
\Sigma_B(z) &=& - \frac{A_c}{2} 
\frac{e^{i\alpha_c\pi/2}}{\sin[\alpha_c\pi/2]} |z|^{-\alpha_c}
\end{eqnarray}
and similarly for $G_c(z)$ and $G_B(z)$. Finally, from Dyson's
equation~(\ref{Gc})-(\ref{GB}), we have
$G_c(z) \sim -1/\Sigma_c(z)$ and $G_B(z) \sim -1/\Sigma_B(z)$, providing
relations between amplitudes $A_c$, $A_B$ and exponents $\alpha_c$, $\alpha_B$. 
After some manipulations, we find:
\begin{equation}
\alpha_B = - \alpha_c = \frac{2}{\pi} \arctan\frac{1}{\sqrt{\gamma}}
\end{equation}

\bibliographystyle{apsrev}

\begin{thebibliography}{32}
\expandafter\ifx\csname natexlab\endcsname\relax\def\natexlab#1{#1}\fi
\expandafter\ifx\csname bibnamefont\endcsname\relax
  \def\bibnamefont#1{#1}\fi
\expandafter\ifx\csname bibfnamefont\endcsname\relax
  \def\bibfnamefont#1{#1}\fi
\expandafter\ifx\csname citenamefont\endcsname\relax
  \def\citenamefont#1{#1}\fi
\expandafter\ifx\csname url\endcsname\relax
  \def\url#1{\texttt{#1}}\fi
\expandafter\ifx\csname urlprefix\endcsname\relax\def\urlprefix{URL }\fi
\providecommand{\bibinfo}[2]{#2}
\providecommand{\eprint}[2][]{\url{#2}}

\bibitem[{\citenamefont{Hewson}(1996)}]{Hewson}
\bibinfo{author}{\bibfnamefont{A.}~\bibnamefont{Hewson}},
  \emph{\bibinfo{title}{The Kondo problem to heavy fermions}}
  (\bibinfo{publisher}{Cambridge}, \bibinfo{year}{1996}).

\bibitem[{\citenamefont{Georges et~al.}(1996)\citenamefont{Georges, Kotliar,
  Krauth, and Rozenberg}}]{RMP_DMFT}
\bibinfo{author}{\bibfnamefont{A.}~\bibnamefont{Georges}},
  \bibinfo{author}{\bibfnamefont{G.}~\bibnamefont{Kotliar}},
  \bibinfo{author}{\bibfnamefont{W.}~\bibnamefont{Krauth}}, \bibnamefont{and}
  \bibinfo{author}{\bibfnamefont{M.}~\bibnamefont{Rozenberg}},
  \bibinfo{journal}{\rmp} \textbf{\bibinfo{volume}{68}}, \bibinfo{pages}{13}
  (\bibinfo{year}{1996}).

\bibitem[{\citenamefont{Nozi\`eres and Blandin}(1980)}]{Noz_Bland}
\bibinfo{author}{\bibfnamefont{P.}~\bibnamefont{Nozi\`eres}} \bibnamefont{and}
  \bibinfo{author}{\bibfnamefont{A.}~\bibnamefont{Blandin}},
  \bibinfo{journal}{J. Physique} \textbf{\bibinfo{volume}{41}},
  \bibinfo{pages}{193} (\bibinfo{year}{1980}).

\bibitem[{\citenamefont{Bulla and Vojta}(2003)}]{Bulla_Vojta}
\bibinfo{author}{\bibfnamefont{R.}~\bibnamefont{Bulla}} \bibnamefont{and}
  \bibinfo{author}{\bibfnamefont{M.}~\bibnamefont{Vojta}}, \bibinfo{journal}{in
  {\it Concepts in Electron Correlations}, A.C. Hewson and V. Zlatic (eds.),
  Kluwer Academic Publishers, Dordrecht}  (\bibinfo{year}{2003}), \eprint{{\tt
  cond-mat/0210015}}.

\bibitem[{\citenamefont{Lee et~al.}(1986)\citenamefont{Lee, Rice, Serene, Sham,
  and Wilkins}}]{revue_HF}
\bibinfo{author}{\bibfnamefont{P.~A.} \bibnamefont{Lee}},
  \bibinfo{author}{\bibfnamefont{T.~M.} \bibnamefont{Rice}},
  \bibinfo{author}{\bibfnamefont{J.~W.} \bibnamefont{Serene}},
  \bibinfo{author}{\bibfnamefont{L.~J.} \bibnamefont{Sham}}, \bibnamefont{and}
  \bibinfo{author}{\bibfnamefont{J.~W.} \bibnamefont{Wilkins}},
  \bibinfo{journal}{Comment. Cond. Mat. Phys.} \textbf{\bibinfo{volume}{12}},
  \bibinfo{pages}{99} (\bibinfo{year}{1986}).

\bibitem[{\citenamefont{Stewart}(2001)}]{Stewart}
\bibinfo{author}{\bibfnamefont{G.~R.} \bibnamefont{Stewart}},
  \bibinfo{journal}{\rmp} \textbf{\bibinfo{volume}{73}}, \bibinfo{pages}{797}
  (\bibinfo{year}{2001}).

\bibitem[{\citenamefont{Schr\"oder et~al.}(2000)\citenamefont{Schr\"oder,
  Aeppli, Coldea, Adams, Stockert, von L\"ohneysen, Bucher, Ramazashvili, and
  Coleman}}]{schroeder}
\bibinfo{author}{\bibfnamefont{A.}~\bibnamefont{Schr\"oder}},
  \bibinfo{author}{\bibfnamefont{G.}~\bibnamefont{Aeppli}},
  \bibinfo{author}{\bibfnamefont{R.}~\bibnamefont{Coldea}},
  \bibinfo{author}{\bibfnamefont{M.}~\bibnamefont{Adams}},
  \bibinfo{author}{\bibfnamefont{O.}~\bibnamefont{Stockert}},
  \bibinfo{author}{\bibfnamefont{H.}~\bibnamefont{von L\"ohneysen}},
  \bibinfo{author}{\bibfnamefont{E.}~\bibnamefont{Bucher}},
  \bibinfo{author}{\bibfnamefont{R.}~\bibnamefont{Ramazashvili}},
  \bibnamefont{and} \bibinfo{author}{\bibfnamefont{P.}~\bibnamefont{Coleman}},
  \bibinfo{journal}{Nature} \textbf{\bibinfo{volume}{407}},
  \bibinfo{pages}{351} (\bibinfo{year}{2000}).

\bibitem[{\citenamefont{Kouwenhoven and Glazman}(2001)}]{revival_kondo}
\bibinfo{author}{\bibfnamefont{L.}~\bibnamefont{Kouwenhoven}} \bibnamefont{and}
  \bibinfo{author}{\bibfnamefont{L.}~\bibnamefont{Glazman}},
  \bibinfo{journal}{Physics World} \textbf{\bibinfo{volume}{14}},
  \bibinfo{pages}{33} (\bibinfo{year}{2001}).

\bibitem[{\citenamefont{Oreg and Goldhaber-Gordon}(2003)}]{Oreg_GG}
\bibinfo{author}{\bibfnamefont{Y.}~\bibnamefont{Oreg}} \bibnamefont{and}
  \bibinfo{author}{\bibfnamefont{D.}~\bibnamefont{Goldhaber-Gordon}},
  \bibinfo{journal}{\prl} \textbf{\bibinfo{volume}{90}},
  \bibinfo{pages}{136602} (\bibinfo{year}{2003}).

\bibitem[{\citenamefont{Florens and Rosch}(2004)}]{SF_AR}
\bibinfo{author}{\bibfnamefont{S.}~\bibnamefont{Florens}} \bibnamefont{and}
  \bibinfo{author}{\bibfnamefont{A.}~\bibnamefont{Rosch}},
  \bibinfo{journal}{\prl} \textbf{\bibinfo{volume}{92}},
  \bibinfo{pages}{216601} (\bibinfo{year}{2004}).

\bibitem[{\citenamefont{Anders et~al.}(2003)\citenamefont{Anders, Lebanon, and
  Schiller}}]{anders}
\bibinfo{author}{\bibfnamefont{F.~B.} \bibnamefont{Anders}},
  \bibinfo{author}{\bibfnamefont{E.}~\bibnamefont{Lebanon}}, \bibnamefont{and}
  \bibinfo{author}{\bibfnamefont{A.}~\bibnamefont{Schiller}}
  (\bibinfo{year}{2003}), \eprint{{\tt cond-mat/0311502}}.

\bibitem[{\citenamefont{Coleman and P\'epin}(2003)}]{Coleman_Pepin}
\bibinfo{author}{\bibfnamefont{P.}~\bibnamefont{Coleman}} \bibnamefont{and}
  \bibinfo{author}{\bibfnamefont{C.}~\bibnamefont{P\'epin}},
  \bibinfo{journal}{\prb} \textbf{\bibinfo{volume}{68}},
  \bibinfo{pages}{220405} (\bibinfo{year}{2003}).

\bibitem[{\citenamefont{Giamarchi et~al.}(1993)\citenamefont{Giamarchi, Varma,
  Ruckenstein, and Nozi\`eres}}]{Giam_Varma}
\bibinfo{author}{\bibfnamefont{T.}~\bibnamefont{Giamarchi}},
  \bibinfo{author}{\bibfnamefont{C.~M.} \bibnamefont{Varma}},
  \bibinfo{author}{\bibfnamefont{A.~E.} \bibnamefont{Ruckenstein}},
  \bibnamefont{and}
  \bibinfo{author}{\bibfnamefont{P.}~\bibnamefont{Nozi\`eres}},
  \bibinfo{journal}{\prl} \textbf{\bibinfo{volume}{70}}, \bibinfo{pages}{3967}
  (\bibinfo{year}{1993}).

\bibitem[{\citenamefont{Schlottmann}(2000)}]{UKM_BA}
\bibinfo{author}{\bibfnamefont{P.}~\bibnamefont{Schlottmann}},
  \bibinfo{journal}{\prl} \textbf{\bibinfo{volume}{84}}, \bibinfo{pages}{1559}
  (\bibinfo{year}{2000}).

\bibitem[{\citenamefont{Florens}(2004)}]{SF_scaling}
\bibinfo{author}{\bibfnamefont{S.}~\bibnamefont{Florens}},
  \bibinfo{journal}{\prb} \textbf{\bibinfo{volume}{69}},
  \bibinfo{pages}{113103} (\bibinfo{year}{2004}).

\bibitem[{\citenamefont{Coleman}(1987)}]{Coleman2}
\bibinfo{author}{\bibfnamefont{P.}~\bibnamefont{Coleman}},
  \bibinfo{journal}{\prb} \textbf{\bibinfo{volume}{35}}, \bibinfo{pages}{5072}
  (\bibinfo{year}{1987}).

\bibitem[{\citenamefont{Cox and Ruckenstein}(1993)}]{Cox_Ruck_multi}
\bibinfo{author}{\bibfnamefont{D.~L.} \bibnamefont{Cox}} \bibnamefont{and}
  \bibinfo{author}{\bibfnamefont{A.~E.} \bibnamefont{Ruckenstein}},
  \bibinfo{journal}{\prl} \textbf{\bibinfo{volume}{71}}, \bibinfo{pages}{1613}
  (\bibinfo{year}{1993}).

\bibitem[{\citenamefont{Parcollet et~al.}(1998)\citenamefont{Parcollet,
  Georges, Kotliar, and Sengupta}}]{OP_AG_multi2}
\bibinfo{author}{\bibfnamefont{O.}~\bibnamefont{Parcollet}},
  \bibinfo{author}{\bibfnamefont{A.}~\bibnamefont{Georges}},
  \bibinfo{author}{\bibfnamefont{G.}~\bibnamefont{Kotliar}}, \bibnamefont{and}
  \bibinfo{author}{\bibfnamefont{A.}~\bibnamefont{Sengupta}},
  \bibinfo{journal}{\prb} \textbf{\bibinfo{volume}{58}}, \bibinfo{pages}{3794}
  (\bibinfo{year}{1998}).

\bibitem[{\citenamefont{Parcollet and Georges}(1997)}]{OP_AG_multi1}
\bibinfo{author}{\bibfnamefont{O.}~\bibnamefont{Parcollet}} \bibnamefont{and}
  \bibinfo{author}{\bibfnamefont{A.}~\bibnamefont{Georges}},
  \bibinfo{journal}{\prl} \textbf{\bibinfo{volume}{79}}, \bibinfo{pages}{4665}
  (\bibinfo{year}{1997}).

\bibitem[{\citenamefont{Georges}()}]{AG_priv}
\bibinfo{author}{\bibfnamefont{A.}~\bibnamefont{Georges}},
  \bibinfo{note}{private communication}.

\bibitem[{\citenamefont{Bickers}(1987)}]{bickers_RMP}
\bibinfo{author}{\bibfnamefont{N.~E.} \bibnamefont{Bickers}},
  \bibinfo{journal}{\rmp} \textbf{\bibinfo{volume}{59}}, \bibinfo{pages}{845}
  (\bibinfo{year}{1987}).

\bibitem[{\citenamefont{Auerbach and Levin}(1986)}]{Auerbach_Levin}
\bibinfo{author}{\bibfnamefont{A.}~\bibnamefont{Auerbach}} \bibnamefont{and}
  \bibinfo{author}{\bibfnamefont{K.}~\bibnamefont{Levin}},
  \bibinfo{journal}{\prl} \textbf{\bibinfo{volume}{57}}, \bibinfo{pages}{877}
  (\bibinfo{year}{1986}).

\bibitem[{\citenamefont{Kiselev et~al.}(2001)\citenamefont{Kiselev, Feldmann,
  and Oppermann}}]{kiselev}
\bibinfo{author}{\bibfnamefont{M.~N.} \bibnamefont{Kiselev}},
  \bibinfo{author}{\bibfnamefont{H.}~\bibnamefont{Feldmann}}, \bibnamefont{and}
  \bibinfo{author}{\bibfnamefont{R.}~\bibnamefont{Oppermann}},
  \bibinfo{journal}{Eur. Phys. J B} \textbf{\bibinfo{volume}{22}},
  \bibinfo{pages}{53} (\bibinfo{year}{2001}).

\bibitem[{\citenamefont{Mueller-Hartmann}(1984)}]{mueller_hart}
\bibinfo{author}{\bibfnamefont{E.}~\bibnamefont{Mueller-Hartmann}},
  \bibinfo{journal}{Z. Phys. B} \textbf{\bibinfo{volume}{57}},
  \bibinfo{pages}{281} (\bibinfo{year}{1984}).

\bibitem[{\citenamefont{Parcollet}()}]{OP_these}
\bibinfo{author}{\bibfnamefont{O.}~\bibnamefont{Parcollet}}, \bibinfo{note}{ph.
  D Thesis, Paris VI University}.

\bibitem[{\citenamefont{Florens}()}]{SF_todo}
\bibinfo{author}{\bibfnamefont{S.}~\bibnamefont{Florens}}, \bibinfo{note}{in
  preparation}.

\bibitem[{\citenamefont{Zinn-Justin and Andrei}(1998)}]{PZJ}
\bibinfo{author}{\bibfnamefont{P.}~\bibnamefont{Zinn-Justin}} \bibnamefont{and}
  \bibinfo{author}{\bibfnamefont{N.}~\bibnamefont{Andrei}},
  \bibinfo{journal}{Nucl. Phys. B} \textbf{\bibinfo{volume}{B528}},
  \bibinfo{pages}{648} (\bibinfo{year}{1998}).

\bibitem[{\citenamefont{LeHur}(1999)}]{KLH_UKM}
\bibinfo{author}{\bibfnamefont{K.}~\bibnamefont{LeHur}},
  \bibinfo{journal}{\prl} \textbf{\bibinfo{volume}{83}}, \bibinfo{pages}{848}
  (\bibinfo{year}{1999}).

\bibitem[{\citenamefont{Pustilnik and Glazman}(2004)}]{S1_kondo_dot}
\bibinfo{author}{\bibfnamefont{M.}~\bibnamefont{Pustilnik}} \bibnamefont{and}
  \bibinfo{author}{\bibfnamefont{L.}~\bibnamefont{Glazman}},
  \bibinfo{journal}{J. Phys: Condens. Matter} \textbf{\bibinfo{volume}{16}},
  \bibinfo{pages}{R513} (\bibinfo{year}{2004}).

\bibitem[{\citenamefont{Sengupta et~al.}(2004)\citenamefont{Sengupta,
  Sampathkumaran, Rayaprol, Nakano, Hedo, Abliz, Fujiwara, Uwatoko, Shigetoh,
  Takabatake et~al.}}]{sampath}
\bibinfo{author}{\bibfnamefont{K.}~\bibnamefont{Sengupta}},
  \bibinfo{author}{\bibfnamefont{E.~V.} \bibnamefont{Sampathkumaran}},
  \bibinfo{author}{\bibfnamefont{S.}~\bibnamefont{Rayaprol}},
  \bibinfo{author}{\bibfnamefont{T.}~\bibnamefont{Nakano}},
  \bibinfo{author}{\bibfnamefont{M.}~\bibnamefont{Hedo}},
  \bibinfo{author}{\bibfnamefont{M.}~\bibnamefont{Abliz}},
  \bibinfo{author}{\bibfnamefont{N.}~\bibnamefont{Fujiwara}},
  \bibinfo{author}{\bibfnamefont{Y.}~\bibnamefont{Uwatoko}},
  \bibinfo{author}{\bibfnamefont{K.}~\bibnamefont{Shigetoh}},
  \bibinfo{author}{\bibfnamefont{T.}~\bibnamefont{Takabatake}},
  \bibnamefont{et~al.}, \bibinfo{journal}{\prb} \textbf{\bibinfo{volume}{70}},
  \bibinfo{pages}{064406} (\bibinfo{year}{2004}).

\bibitem[{\citenamefont{Coleman and Paul}(2004)}]{Coleman_Paul}
\bibinfo{author}{\bibfnamefont{P.}~\bibnamefont{Coleman}} \bibnamefont{and}
  \bibinfo{author}{\bibfnamefont{I.}~\bibnamefont{Paul}}
  (\bibinfo{year}{2004}), \eprint{{\tt cond-mat/0404001}}.

\bibitem[{\citenamefont{Mehta et~al.}(2004)\citenamefont{Mehta, Borda, Zarand,
  Andrei, and Coleman}}]{singular}
\bibinfo{author}{\bibfnamefont{P.}~\bibnamefont{Mehta}},
  \bibinfo{author}{\bibfnamefont{L.}~\bibnamefont{Borda}},
  \bibinfo{author}{\bibfnamefont{G.}~\bibnamefont{Zarand}},
  \bibinfo{author}{\bibfnamefont{N.}~\bibnamefont{Andrei}}, \bibnamefont{and}
  \bibinfo{author}{\bibfnamefont{P.}~\bibnamefont{Coleman}}
  (\bibinfo{year}{2004}), \eprint{{\tt cond-mat/0404122}}.

\end{thebibliography}

\newcommand{\npb}{Nucl. Phys.}\newcommand{\adv}{Adv.
  Phys.}\newcommand{\epl}{Europhys. Lett.}

\end{document}